\documentclass[prb,twocolumn,showpacs,amsmath,amssymb]{revtex4}

\usepackage{graphicx}
\usepackage{epstopdf}
\usepackage{bm}
\usepackage{gensymb}
\usepackage{floatrow}
\usepackage{placeins}

\DeclareGraphicsExtensions{.eps,.jpg,.pdf}

\begin{document}

\title{Very large magnetoresistance in Fe$_{0.28}$TaS$_2$ single crystals}
\author{Will J. Hardy,$^1$ Chih-Wei Chen,$^1$ A. Marcinkova,$^1$ Heng Ji,$^{1}$ Jairo Sinova,$^2$ D. Natelson$^1$ and E. Morosan$^1$}
\address{$^1$Department of Physics and Astronomy, Rice University, Houston, TX 77005 USA}
\address{$^2$Institute of Physics, Johannes Gutenberg Universit{\"a}t, Mainz, 55128 Germany}
\date{\today}

\begin{abstract}

Magnetic moments intercalated into layered transition metal dichalcogenides are an excellent system for investigating the rich physics associated with magnetic ordering in a strongly anisotropic, strong spin-orbit coupling environment.  We examine electronic transport and magnetization in Fe$_{0.28}$TaS$_{2}$, a highly anisotropic ferromagnet with a Curie temperature $T_{\mathrm{C}} \sim 68.8$~K.  We find anomalous Hall data confirming a dominance of spin-orbit coupling in the magnetotransport properties of this material, and a remarkably large field-perpendicular-to-plane MR exceeding 60\% at 2~K, much larger than the typical MR for bulk metals, and comparable to state-of-the-art GMR in thin film heterostructures, and smaller only than CMR in Mn perovskites or high mobility semiconductors. Even within the Fe$_x$TaS$_2$  series, for the current $x$ = 0.28 single crystals the MR is nearly $100\times$ higher than that found previously in the commensurate compound Fe$_{0.25}$TaS$_{2}$.  After considering alternatives, we argue that the large MR arises from spin disorder scattering in the strong spin-orbit coupling environment, and suggest that this can be a design principle for materials with large MR.

\end{abstract}


\maketitle

\section{Introduction}

Spintronics, which concerns the effects on transport due to the coupled spin and charge degrees of freedom of the electron, has raised intense interest due to its broad industrial applications and theoretical challenges.\cite{Wolf_2001, Zutic_2004, Awschalom_2007, Chappert_2007, Bader_2010} These magnetic transport properties underlie giant, tunneling, and colossal magnetoresistance (GMR, TMR, and CMR),\cite{Baibich_1988,Binasch_1989,Miyazaki_1995,Moodera_1995, Jin_1994} tunneling anisotropic magnetoresistance (TAMR),\cite{Gould_2004, Saito_2005} and the anomalous Hall effect (AHE).\cite{Toyosaki_2004} Both GMR and TMR are widely observed in thin films\cite{Baibich_1988,Binasch_1989,Miyazaki_1995,Moodera_1995} where the magnetic coupling between layers can be artificially tuned. Observation in bulk materials\cite{Kusters_1989, Tokura_1999} revealed that CMR can be a bulk material property. Many mechanisms were suggested for the large magnetoresistance (MR) observed in bulk materials: nanoscale phase separation of the metallic ferromagnetic and insulating antiferromagnetic clusters in manganites;\cite{Uehara_1999, Dagotto_2001} metamagnetic transitions in rare earth intermetallics;\cite{Sechovsky_1994, Janssen_1997} and metal-insulator transitions and double exchange interactions for transition metal oxides.\cite{Hwang_1995, Millis_1996, Hammer_2004, Nakamura_2009}

While structures that exhibit GMR and TMR are already widely used in electronic devices, there remains strong technological and fundamental interest in homogeneous materials that exhibit large magnetoresistive effects. Moreover, since ordinary MR effects in bulk metals are typically only a few percent, understanding any occurrences of enhanced MR effects in bulk is of fundamental interest.  In the ongoing search for new magnetic materials, transition metal dichalcogenides (TMDs) may be ideal candidates, due to their layered crystal structure and ease of intercalation with magnetic elements.\cite{Wilson1969, Wilson1975, Parkin_1980_1, Parkin_1980_2} For nearly forty years, the family of layered compounds Fe$_x$TaS$_2$ has been the subject of sustained inquiry focused on a surprising variety of anisotropic ferromagnetic properties.\cite{Eibschutz_1975, Morosan_2007} Prior studies have demonstrated that tuning the Fe concentration allows control of these magnetic properties, and measurements of magnetization, MR, and the anomalous Hall effect have been effective probes of the resulting modifications in behavior.\cite{Eibschutz_1975,  Eibschutz_1981, Dijkstra_1989, Narita_1994, Morosan_2007, Checkelsky_2008} Here, we report experimental characterization of such a compound, with $x~\approx$ 0.28, which exhibits MR in the ordered state exceeding 60\%, nearly two orders of magnitude larger than was previously measured. By comparing our complementary results from bulk and thin exfoliated samples, we conclude that the large observed change in resistance is intrinsic and does not result from size-dependent phenomena, such as domain wall scattering.  We argue that spin disorder scattering in the presence of strong spin-orbit coupling is the mechanism behind this MR, and that this is a potential paradigm for creating homogeneous materials with large MR.  These observations suggest that the TMDs are rich targets for further theoretical study and potential industrial applications.\cite{Xu_2014} 

\section{Methods}

Single crystals of Fe$_{0.28}$TaS$_2$ were prepared using iodine vapor transport in a sealed quartz tube, as described elsewhere.\cite{Morosan_2007} The typical size of the resulting bulk Fe$_{0.28}$TaS$_2$ single crystals was 2$\times$2$\times$0.1 mm$^3$. Powder x-ray diffraction revealed the expected Fe$_{0.28}$TaS$_2$ phase, with the lattice parameters consistent with a composition $x$ between 0.20 and 0.34.\cite{Eibschutz_1981} Energy-dispersive spectroscopy (EDS) and inductively coupled plasma (ICP) on bulk samples as well were used to more precisely determine the Fe concentration to be $x~=$ 0.28$\pm 3 \%$. EDS data were collected using a scanning electron microscope (SEM) equipped with an energy-dispersive spectroscopy (EDS) detector. ICP data were collected using a Perkin Elmer Optima 8300 ICP-OES system. The iron concentration of the sample was derived by comparison with commercial iron pure single-element standards (Perkin Elmer). Selected area electron diffraction (SAED) was also performed at room temperature on a bulk single crystal, ground in ethanol and placed on a holey carbon TEM grid.

The exfoliated samples were prepared using the tape exfoliation method.\cite{Novoselov_2004} Bulk Fe$_{0.28}$TaS$_2$ single crystals were mechanically cleaved using blue Nitto SPV 224 tape, and the resulting exfoliated crystals were deposited onto an oxidized silicon wafer (300 nm or 2 $\mu$m oxide thickness). Metallic contacts were defined using standard electron beam lithography and development. Contact metals were then deposited by electron beam evaporation of a Ti, Cr, or Fe adhesion layer ($\sim$ 3 nm) and Au (50 nm); an extra 20 nm of Au was added by sputtering. 

For the exfoliated samples, the thickness was determined using atomic force microscopy (AFM). The measured thickness, with average values between 80 and 180 nm, varied by up to 21$\%$ within each sample.  Scanning electron microscopy (SEM) images showed that the exfoliated flakes had lateral dimensions on the order of 10 $\mu$m, with variation from sample to sample.  Thinner samples could only be produced with lateral dimensions much smaller than 10 $\mu$m due to relatively strong bonding between the layers compared to, \textit{e.g.}, graphite. Two exfoliated samples were prepared with electrodes configured to enable Hall measurements as well as conventional MR, while a third exfoliated sample was prepared for MR alone.  Voltage probes were separated by less than 5~$\mu$m in these devices.

\begin{figure}[b!]
\includegraphics[width=1.0\columnwidth,clip]{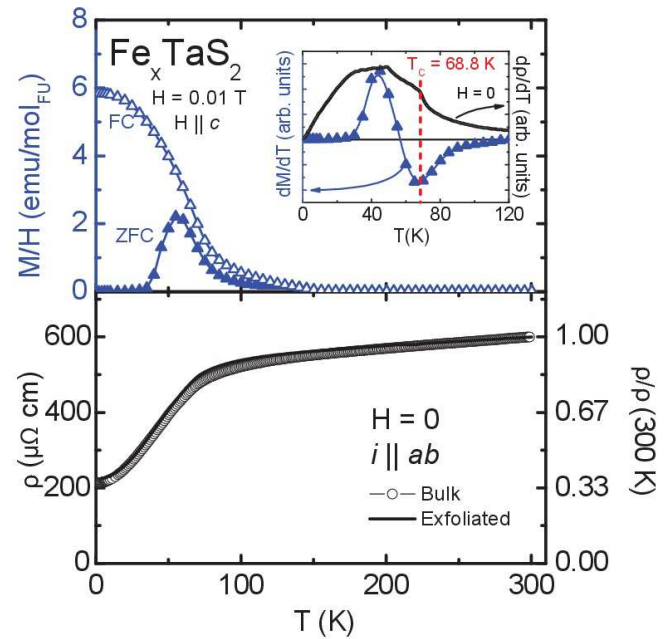}
\caption{\label{magnetization_and_resistivity} (a) ZFC (solid symbols) and FC (open symbols) temperature-dependent magnetic susceptibility of a bulk sample measure in an applied field H = 0.1 T, H $\parallel$ c. Inset: The Curie temperature $T_C$ is determined from the minimum in $dM/dT$ (solid symbols) and an inflexion point in $d\rho/dT$ (line).and. (b) Temperature-dependent resistivity of both bulk (open symbols) and exfoliated (solid line) samples.}
\end{figure}

Temperature- and field-dependent magnetization data for bulk Fe$_{x}$TaS$_2$ were collected in a Quantum Design (QD) Magnetic Property Measurement System (MPMS). Temperature- and magnetic field-dependent AC resistivity measurements for both bulk and exfoliated Fe$_{0.28}$TaS$_2$ were performed in a QD Physical Property Measurement System (PPMS) using standard four-probe methods. Additional Hall resistivity data were collected using a five probe configuration for both the bulk and the exfoliated samples. Angle-dependent transport measurements were performed on an exfoliated sample mounted on a QD horizontal rotator insert, which allowed the sample to be rotated relative to the magnetic field direction.

\section{Results and discussion}
\begin{figure}[b!]
\includegraphics[width=1.0\columnwidth,clip]{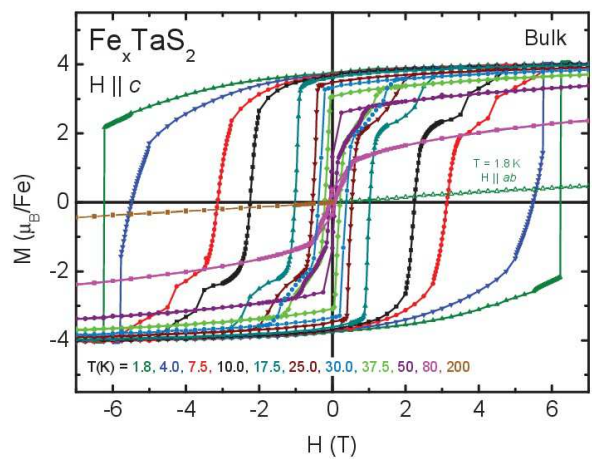}
\caption{\label{magnetization_at_temperatures} H $\parallel~c$ (full symbols) field-dependent magnetization $M(H)$ data at various temperatures, together with the $T$ = 1.8~K, $H \parallel ab$  (open symbols) isotherm. For clarity, the two close isotherms ($H \parallel c$  for $T$ = 200~K and $H \parallel ab$ for $T$ = 1.8~K) are only shown for $H <$ 0 and $H >$ 0, respectively. }
\end{figure}

\begin{figure*}
\includegraphics[width=1.0\columnwidth,clip]{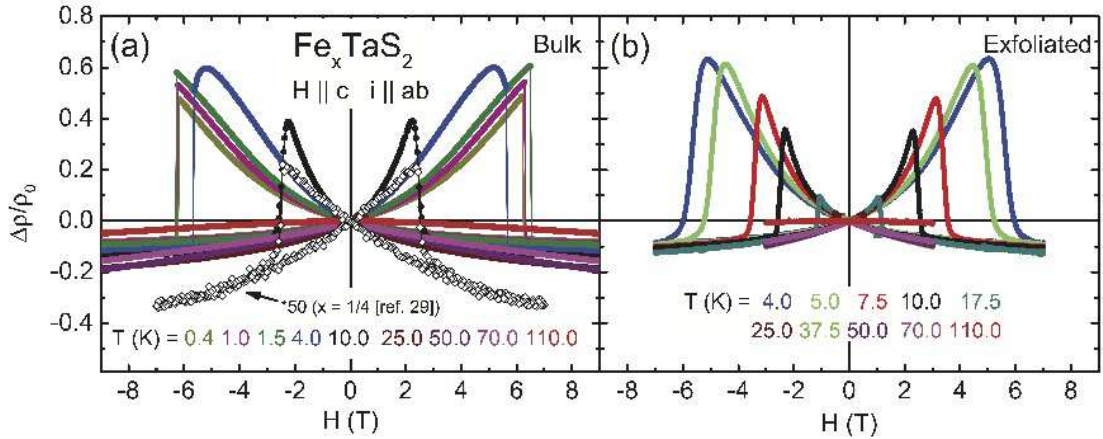}
\caption{\label{GMR} MR of (a) bulk and (b) exfoliated samples at selected temperatures for $H \parallel c$, and the current $i$ $\parallel$ $ab$. }
\end{figure*}

Fe$_x$TaS$_2$ is a unique intercalated transition metal dichalcogenide (TMD), with its strong and non-monotonic dependence of the magnetic properties (the ground state - ferromagnetic or antiferromagnetic, the ordering temperature) on the Fe concentration $x$.\cite{Eibschutz_1981, Dijkstra_1989,Narita_1994} It has been shown that a 3$\%$ difference in the Fe concentration (from $0.25$ to $0.28$) causes a modification of $T_C$ as large as 90 K (from 160 K to 70 K),\cite{Morosan_2007, Dijkstra_1989} {while increasing $x$ from $x~<~0.40$ to $x~\geq~0.40$\cite{Dijkstra_1989,Narita_1994} results in a change of the magnetic interactions from ferro- (FM) to antiferromagnetic (AFM). In the current Fe$_{0.28}$TaS$_2$ single crystals, the $H \parallel c$ temperature-dependent magnetic susceptibility measurements (Fig.~\ref{magnetization_and_resistivity}a) are consistent with the onset of FM order below $\sim$ 70 K upon cooling. The H = 0 temperature dependent resitivity data $\rho(T)$ on bulk (open symbols) and exfoliated (solid line) samples are virtually identical, as can be seen in Fig.~\ref{magnetization_and_resistivity}b. The weakly linear decrease in $\rho(T)$ at high T is indicative of the poor metal behavior in both bulk and exfoliated samples, while a drop below 70 K is consistent with loss of spin disorder scattering in the FM state. The derivatives of the ZFC magnetization data $dM/dT$ (symbols, inset) and the bulk resistivity data $d\rho/dT$ (line, inset) suggest that the Curie temperature T$_C$ is close to 68.8 K, if $T_C$ is determined from the minimum in $dM/dT$ and the inflection point in $d\rho/dT$ (vertical dashed line). The $T_C$ value is consistent with the reported $T_C$ for Fe$_{0.28}$TaS$_2$.\cite{Dijkstra_1989} We do find the onset of irreversibility in the zero-field-cooled (ZFC, solid symbols) and field-cooled (FC, open symbols) $M(T)$ data occurs around 150 K, well above T$_C$ for $x$ = 0.28 and very close to that for $x$ =  0.25.\cite{Morosan_2007} {This may be due to a small amount of Fe ions forming a commensurate superstructure as in Fe$_{0.25}$TaS$_2$, which, however, has very little effect on the transport properties where the transition is not even visible.}}

\begin{figure*}
\includegraphics[width=1.0\columnwidth,clip]{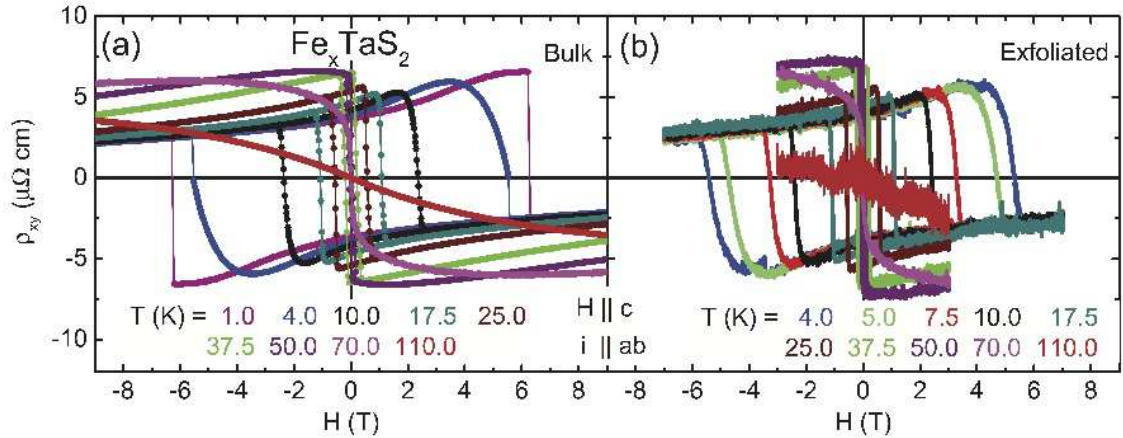}
\caption{\label{AHE} Anomalous Hall resistivity for (a) bulk and (b) exfoliated samples at selected temperatures for $H$ $\parallel$ $c$, and the current $i$ $\parallel$ $ab$.}
\end{figure*}

Remarkable behavior is observed in field-dependent magnetization and resistivity measurements with the magnetic field H along the reported easy axis H$\parallel$ $c$.\cite{Morosan_2007} The magnetization isotherms $M(H)$ of the bulk single crystals (Fig.~\ref{magnetization_at_temperatures}) reveal a sharp switching, similar to that for both Fe$_{0.28}$TaS$_{2}$\cite{Eibschutz_1981} and Fe$_{0.25}$TaS$_{2}$ compounds.\cite{Morosan_2007} The switching field $H_S$ is defined as the field where the magnetization crosses zero and where, as will be shown, the MR $\Delta \rho/ \rho_0$ and Hall resistivivity $\rho_{xy}$ display rapid changes as a function of H $\parallel$ c. In this study, both $H_S$ and the sharpness of the transition decrease with increasing temperature. $H_S$ at 1.8 K has the highest value of 6.23 T, while at T = 4 K, H$_S$ = 5.5 T, very close to value reported for Fe$_{0.28}$TaS$_{2}$.\cite{Eibschutz_1975} A second step-like feature in $M(H)$ appears for 7.5 $\leq~T~\leq$ 80 K and disappears when $T~>$ 200 K. While this could simply be attributed to the small amount of Fe$_x$TaS$_2$ phase with 0.25 $\leq$ x $\leq$ 0.28, this scenario is inconsistent with the absence of the additional M(H) step at the lowest temperatures. Another possible explanation for the second step-like feature could be heat release during the dynamic switching process in the bulk crystals, which could alter the shape of M(H). We do note that the magnetic and transport measurements are reproducible after the samples remain at low temperatures for long periods of time, and after performing multiple field sweeps at different sweep rates. Moreover, the $H \parallel c$ resistivity data $\rho$(H) in Fig.\ref{GMR} and anomalous Hall resistivity in Fig.\ref{AHE} feature a sharp jump at $H_S$. 

MR is a crucial measurement for inferring information about the interactions between itinerant charge carriers and the magnetic degrees of freedom in a variety of magnetic materials.\cite{McGuire_1975, PhysRevB.57.R2037} The MR is defined as: 
\[
\frac{\Delta \rho}{\rho_0} = \frac{\rho_{xx}(H)-\rho_{xx}(0)}{\rho_{xx}(0)} 
\]
where $\rho_{xx}(H)$ is the value of the resistivity in a magnetic field $H$. The $\Delta \rho/ \rho_0$ measurements, with magnetic field H applied along the $c$ axis, were performed at selected temperatures for both bulk and exfoliated Fe$_{0.28}$TaS$_2$ single crystals (Fig.~\ref{GMR}a and b respectively). Below $T_C~\approx~68.8$ K, as the magnetic field H increases from 0 to 9 T, $\Delta \rho/ \rho_0$ smoothly increases to its maximum value at $H_S$ and sharply drops in a very narrow $H$ interval $\Delta H$, followed by a nearly linear decrease up to the maximum measured field $H$ = 9 T. When the magnetic field direction was reversed, the same change in $\Delta \rho/ \rho_0$ was observed, resulting in a bow-tie shape of $\Delta \rho/ \rho_0$ after one full cycle of field sweeping. 

Qualitatively, this MR field-dependence resembles that for Fe$_{0.25}$TaS$_2$.\cite{Morosan_2007} However, the absolute $\Delta \rho/ \rho_0$ values are remarkably high in Fe$_{0.28}$TaS$_2$ (full symbols, Fig. \ref{GMR}a), nearly two orders of magnitude larger than that observed for x = 0.25 (open symbols, Fig. \ref{GMR}a). In both bulk and exfoliated Fe$_{0.28}$TaS$_2$ crystals, the largest $\Delta \rho/ \rho_0$ close to 60$\%$ was observed at T = 4 K (blue, Fig. \ref{GMR}). Furthermore, both  $\Delta \rho/ \rho_0$ and $H_S$ decreased with increasing temperature, and the bow-tie shape of the $\Delta \rho/ \rho_0$ curves disappears above $T_C$ when $\Delta \rho /\rho_{0}$ becomes nearly linear for the whole measured field range. It should be noted that $\Delta H$ is much smaller in bulk ($\sim$ 0.04 T) than in the exfoliated sample ($\sim$ 0.8 T) at lower temperatures, and becomes comparable ($\sim$ 0.3 T) in both as the temperature exceeds 10 K. The broadening of the transition with increasing T in the bulk seems natural, while the opposite effect (sharpening) in the exfoliated sample emphasizes the role of the long range interplanar coupling in Fe$_{0.28}$TaS$_2$. This may imply that long range coupling exists between the Fe ions in different layers, which is weakened in the exfoliated sample, even when 100 nm thick. 

The observed magnitude of the MR in Fe$_{0.28}$TaS$_{2}$, comparable to that seen in GMR and TMR systems, is remarkably large for a homogeneous bulk material not going through a phase transition (as in CMR systems).  A useful point of comparison is (Ga,Mn)As, which has a similar $\rho$ vs $T$ response.\cite{PhysRevB.57.R2037}  This latter material exhibits ordinary AMR, a spin-orbit coupling effect,\cite{McGuire_1975} which is typically at most a few percent in bulk materials based on 3$d$ transition metals. To gain insight into the very large MR in Fe$_{0.28}$TaS$_2$  it is necessary to correlate with other field-dependent measurements, like anomalous Hall effect (AHE).  As previously observed in Fe$_{0.25}$TaS$_2$ and Fe$_{0.28}$TaS$_2$,\cite{Checkelsky_2008, Dijkstra_1989} the Hall resistance $\rho_{xy}$ for both bulk and exfoliated samples displays hysteresis below $T_{C}$, with jumps at $\pm~H_{S}$ (Fig.~\ref{AHE}). As was the case for $\Delta \rho /\rho_{0}$ (Fig. ~\ref{GMR}), $\rho_{xy}$ has a sharper jump at $\pm~H_{S}$ in the bulk sample than in the exfoliated one below 4 K, but then became comparable at higher temperatures. When $H$ exceeds $\pm~H_{S}$, the Hall resistivity $\rho_{xy}$ becomes almost linear in field, a result of the ordinary Hall effect contribution. For temperatures above $T_{C}$, only the ordinary Hall effect is observed, as $\rho_{xy}(H)$ is again nearly linear in $H$.   Note that the Hall coefficient $R_H$ in Fe$_{0.28}$TaS$_2$ does not change sign throughout the ordered state, in contrast to  the situation in Fe$_{0.25}$TaS$_2$.\cite{Checkelsky_2008}  Converting into the Hall conductivity, the change in $\sigma_{xy}$ when passing through $H_{S}$ at 4~K is close to 200~S/cm, essentially the same as that seen in the $x=0.25$ compound,\cite{Checkelsky_2008} and exceeding the values typically seen in (Ga,Mn)As by a factor of five\cite{Nagaosa_2010}.   These results imply that the spin-orbit coupling is very strong in this material and is very similar in the $x=0.28$ and $x=0.25$ compositions. 

We must consider candidate mechanisms to explain the magnetotransport properties of the Fe$_{0.28}$TaS$_2$ single crystals, in particular the remarkably large H $\parallel$ $c$ MR.  One natural possibility is AMR,\cite{McGuire_1975} parametrized in terms of the resistivities measured with the current density $\mathbf{J}$ parallel or perpendicular to the magnetization $\mathbf{M}$, $\rho_{||}$ and $\rho_{\perp}$, respectively.  Generally the difference between the two $\rho_{\Delta} \equiv \rho_{||}-\rho_{\perp}$ is positive. The prior work\cite{Checkelsky_2008} on the $x=0.25$ compound ascribed the small (a maximum $\Delta \rho/\rho_{0} \approx 1.5\%$ at 1.5~K) MR for H $\parallel$ c to a $\rho_{\Delta}$ of +260$~\mu\Omega$-cm and a splaying of the spins as $H \rightarrow H_{S}$ by about 0.1$^{\circ}$. The large value of $\rho_{\Delta}$ is consistent in that case with in-plane MR measurements out to very high fields, showing $\Delta \rho/\rho_{0} \approx 40\%$ for H $\perp$ c and $H = 31$~T, corresponding to a tilting of $M$ away from the $c$ axis by around 15$^{\circ}$.\cite{Checkelsky_2008} Note that in these $x=0.25$ in-plane measurements at 10~K, an in-plane field of several Tesla is able to cant $M$ suffficiently to produce a measured AMR of several percent.

\begin{figure*}
\includegraphics[width=1.0\columnwidth,clip]{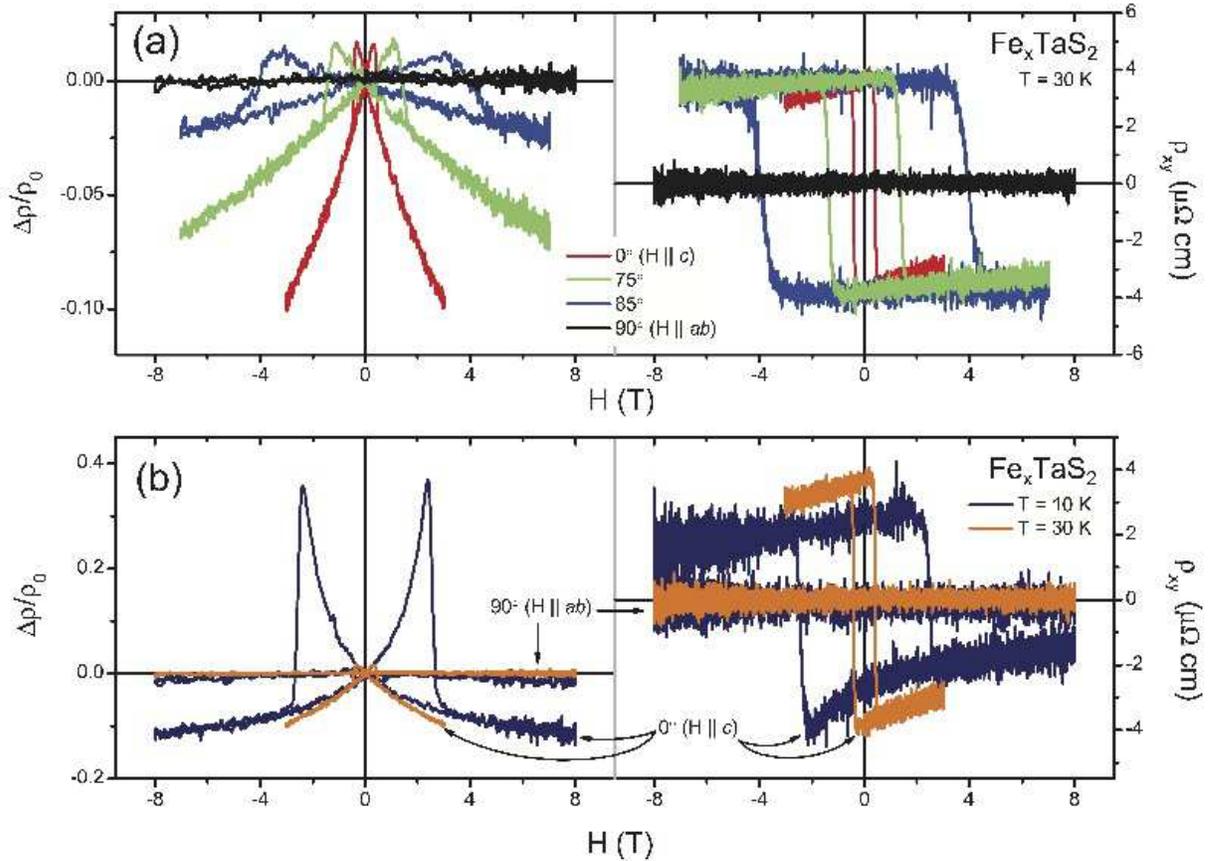}
\caption{\label{AngleDep} Angle-dependent measurements on an exofoliated sample of the longitudinal MR (left) and Hall resistivity (right) as a function of magnetic field H, for H $\parallel$ $c$, and the current $i$ $\parallel$ $ab$.  (a) Data at T = 30 K for various field orientations relative to the $c$ axis. (b) Comparison of $H\parallel$ $c$ and $H\parallel$ $ab$ data for T = 10 and 30 K.}
\end{figure*}

In our $x~=~0.28$ compound, it is not unreasonable to assume a similar magnitude of $\rho_{\Delta}$, given the similarity of the spin-orbit coupling (inferred from the anomalous Hall conductivities) and the switching fields.  Our observed magnitude of $\Delta \rho/\rho_{0}$ for $H\parallel~c$ would then imply a canting or splaying of the spins by tens of degrees immediately prior to magnetization reversal ($|H|~\lesssim~|H_S|$). Indeed, a significant rounding of $M(H)$ (Fig. \ref{magnetization_at_temperatures}) and $\sigma_{xy}(H)$ (Fig. \ref{GMR}) near $H_S$ for $H \parallel~c$ below, \textit{e.g.}, 10 K would at first glance seem to be compatible with this idea. However, angular dependent MR measurements on Fe$_{0.28}$TaS$_2$ strongly disfavor this possibility.  Fig. \ref{AngleDep} displays MR $\Delta \rho/\rho_{0}$ (left) and $\rho_{xy}(H)$ (right) data for (a) different field orientations relative to the $c$-axis and constant temperature $T$ = 30 K, and (b) two extreme field orientations: $H \parallel c$  and $H \parallel  ab$  for $T$ = 10~K (navy) and 30~K (orange).  Within the AMR scenario of canting or splaying of the spins, one would expect significant canting of the magnetization when $H \parallel ab$ if such reorientation of $\mathbf{M}$ could happen with $H$ antialigned to $\mathbf{M}$ along $c$.  Instead, there is almost no detectable magnetoresistive or anomalous Hall response for $H \parallel$ $ab$, and the magnetization response along that field direction (open symbols, Fig. \ref{magnetization_at_temperatures}) is correspondingly weak. This is in contrast to the $x=0.25$ case described above. These observations suggest that the easy axis of magnetization is strongly aligned with the $c$-axis, given that an in-plane field of 8 T is insufficient to produce any detectable MR or Hall signal.  Thus ordinary AMR seems incompatible with the full ensemble of data, and AMR in the $x=0.28$ case appears to be quite different than at $x=0.25$.

\begin{figure}
\includegraphics[width=1.0\columnwidth,clip]{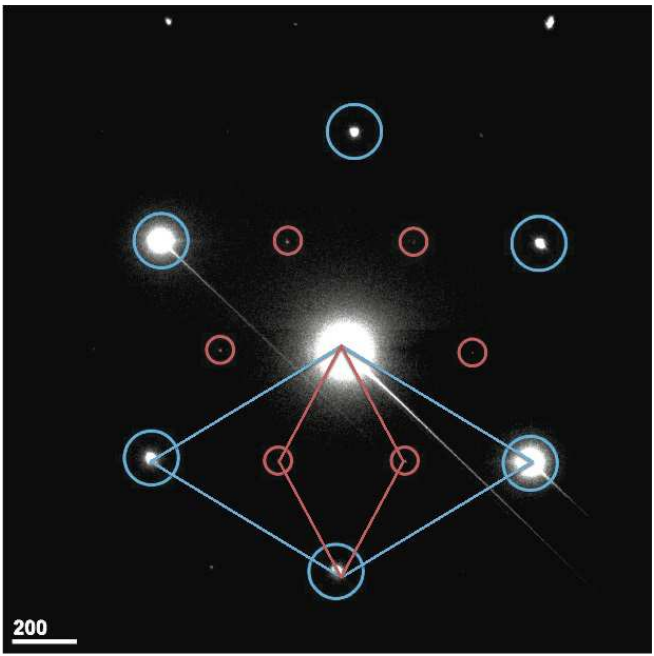}
\caption{\label{SAED} 
SAED pattern of Fe$_{0.28}$TaS$_2$ crystal showing two concentric hexagonal sets of spots:  the main structure (bright, large circles) and superlattice reflections (faint, small circles). The superstructure unit cell (small hexagonal cell) appears rotated by 90$\degree$ from the main structure unit cell (large hexagonal cell).}
\end{figure}

Giant magnetoresistance (GMR)\cite{Baibich_1988,Binasch_1989} is another mechanism capable of producing magnetoresistive effects of tens of percent. GMR results from the interplay between spin-split band structure and the density of states available for each spin species for scattering at the Fermi level. A magnetically inhomogeneous material can exhibit GMR due to current flow between differently aligned magnetic domains.\cite{Xiao_1992} To be a plausible explanation of our data would require that the magnetic domain structure of the material evolve as $H \rightarrow H_{S}$ so that charge transport is forced to take place across an increasingly large number of boundaries between antialigned domains. This can be tested through magneto-optic studies of the domain structure (beyond the scope of the present work). However, the micro-scale exfoliated samples have transport properties that look very similar to those of the bulk crystals, including a lack of any step-like features in the MR or anomalous Hall data as a function of field. This suggests that the flipping of discrete domains near $H_{S}$ and resultant GMR are unlikely to be responsible for the observed large MR. Note further that the domains observed via magneto-optic methods in the $x=0.25$ composition\cite{Vannette_2009} are typically tens of $\mu$m in extent.  In the current x = 0.28 exfoliated samples the few-$\mu$m spacing of the voltage probes combined with the lack of any discrete magnetoresistive or AHE signatures in these devices implies that any domains would have to be much smaller than the $\mu$m scale - very different than the $x=0.25$ case, and difficult to image. Conversely, the similarity in the $M(H)$ data between this study and previous measurements on Fe$_{0.25}$TaS$_2$ suggests that the domain structures are likely very similar.  Therefore, domain wall scattering is unlikely the cause of the large observed MR.

We suggest that the mechanism for the extremely large $H \parallel c$  MR and the near-absence of MR when $H \parallel ab$ is spin disorder scattering.\cite{Haas_1968,Otto_1989} The prominant drop in $\rho(T)$ when $T$ falls below $T_{C}$ is readily apparent in Fig.~\ref{magnetization_and_resistivity}, showing that spin disorder scattering accounts for approximately 50\% of the total scattering relevant to the resistivity above $T_{C}$. In the case of large spin-orbit coupling (SOC, as indicated by the size of the anomalous Hall conductivity in this material), it is not surprising that spin disorder can be so important. Rather than carrier-magnon scattering or Kondo physics, with the strong anisotropy and SOC the proposed mechanism for the large MR in the current x = 0.28 system is scattering from a (quasistatic) disordered arrangement of antialigned moments.  In the presence of strong SOC, such spin disorder can be very effective at scattering carriers relative to ordinary potential disorder, since it mixes spin channels and therefore permits greater phase space for scattering.

When electron diffraction measurements are performed on the Fe$_{0.28}$TaS$_2$ single crystals (Fig.~\ref{SAED}), two concentric sets of spots are observed in the $ab$ plane, each with sixfold symmetry. The bright spots (large circles) are due to the main TaS$_2$ phase, while the faint spots (small circles) are assumed to result from an ordered Fe superstructure. When compared to the diffraction patterns presented in a recent study by Horibe $et~al.$,\cite{Horibe2014} the present SAED pattern appears more similar to that of Fe$_{1/3}$TaS$_2$ than that of Fe$_{1/4}$TaS$_2$, with the interior hexagon rotated by 90\degree in relation to the outer one and the resulting superstructure close to $\sqrt{3}\times\sqrt{3}$.  The appearance of the superstructure spots in the electron diffraction (Fig.~\ref{SAED}) indicates that it may be useful to think about the $x~=~0.28$ system as a compound with a commensurate $x~=~0.25$ Fe structure with additional Fe local moments ($x~=~0.25~+~\delta$), or $x~=~0.33$ Fe structure with missing Fe local moments ($x~=~0.33~-~\delta$) with very small $\delta$ ($\delta~\leq~0.05$). In either case, the moments in a disordered environment, while coupled ferromagnetically to the bulk, would be expected to have weaker exchange interactions\cite{Ko_2011} than those on the superstructure sites, and hence easier to antialign with the field as $(H\parallel c) \rightarrow H_{S}$. The maximum MR for this field orientation is seen at $H_{S}$ as the spins reverse their orientation, leading to an increase in scattering comparable to the spin-disorder contribution to $\rho$. In other words, during the MR measurement, the antialignment of a significant fraction of the local moments as the field strength is increased (antiparallel to the bulk magnetization) results in enhanced scattering and increased resistance. Once the remaining spins flip to become aligned with the external field, spin disorder scattering is greatly reduced, causing a sharp drop in resistance. Canting of the moments is disfavored by the large magnetic anisotropy, while enhanced scattering (relative to potential scattering) is favored due to strong SOC and channel mixing. 

Additional experiments can be used to test this hypothesis. This explanation assumes a population of weakly-coupled,  easier-to-reorient spins due to deviations from the $x~=~0.25$ stoichiometry. One would therefore expect a monotonic increase in the the $H\parallel$ $c$ MR as $x$ is increased from $x~=~0.25$ to $x~=~0.28$. The dynamics of the spin reorientation should also be manifested in the MR response in this case, though no field sweep rate dependence has been observed so far. Optical perturbation of the local moment orientation would also be expected to lead to large resistive effects.

\begin{figure}[h]
\includegraphics[width=1.0\columnwidth,clip]{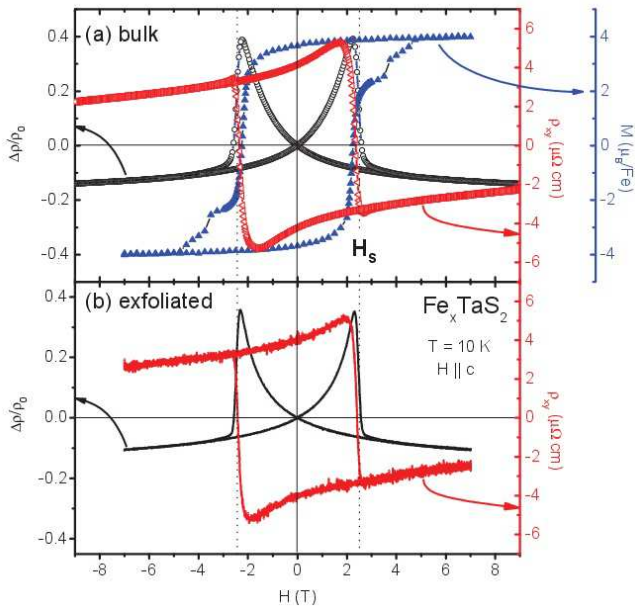}
\caption{\label{determination_Hs} Determination of the switching field $H_{S}$ for (a) bulk and (b) exfoliated samples from $M(H)$ (blue), MR (black) and anomalous Hall resistivity (red). The vertical dashed line marks the switching field $H_S$, as determined from the field values where M(H) and $\rho_{xy}$ cross H = 0, and where the fastest drop in $\Delta\rho/\rho_0$ occured.}
\end{figure}

\begin{figure}[h]
\includegraphics[width=1.0\columnwidth,clip]{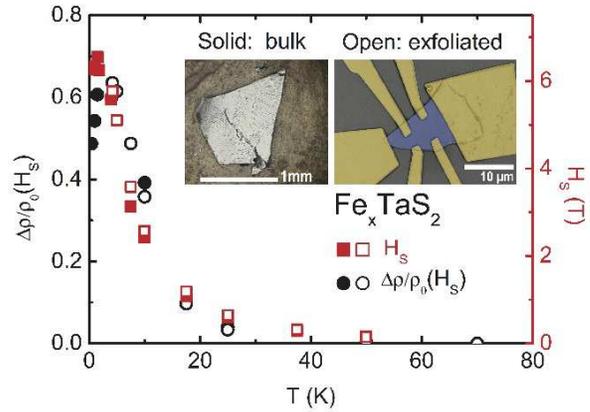}
\caption{\label{Comparison} Comparison of $H_S$ and magnetoresistivity peak height values as a function of temperature for bulk (solid symbols) and exfoliated (open symbols) samples. $H_S$ increased monotonically with decreasing temperature, while the magnetoresistivity peak height increased with decreasing temperature until 4 K, and then decreased at lower temperatures. Inset: Image of a typical bulk sample (left), and false-color SEM image of a typical exfoliated sample with metal contacts (right).}
\end{figure}

In conclusion, we show that Fe$_{0.28}$TaS$_2$ single crystals display remarkably large MR, up to 60$\%$, when the applied magnetic field $H\parallel c$. Both the magnetization and transport properties appear nearly insensitive to sample thickness down to $\sim$ 100 nm, as measurements on bulk and exfoliated single crystals are nearly indistinguishable. As is illustrated in Fig. \ref{determination_Hs} for $T = 10$~K, the switching field $H_S$ values observed from magnetization and magneto-transport measurements on both bulk and exfoliated samples are very close at all temperatures up to $T_{C}$. The resulting temperature dependence of $H_S$ (squares) and $\Delta \rho/\rho_0$ at $H_S$ (circles) shown in Fig. \ref{Comparison} is indeed identical for both the bulk (full symbols) and exfoliated (open symbols) samples. Moreover, the non-monotonic change with x of the ordering temperature $T_C$ and switching field values $H_S$ between the Fe$_{0.28}$TaS$_2$ system and the previously reported Fe$_{0.25}$TaS$_2$ superstructure\cite{Morosan_2007}, and, more significantly, the nearly two order of magnitude enhancement of MR in the former compound, appear to be consistent with a scenario of disordered Fe moments mixed with a Fe superstructure. This scenario is even more plausible, given  the experimental evidence we present to rule out other likely possibilities, such as AMR or an analog of GMR due to domain structure.  The spin disorder scattering scenario reveals a design principle for intrinsically magnetoresistive materials without the need for multilayers or metal-insulator transitions coupled to magnetism.  Conditions favoring maximal MR would include: single crystal materials, so that grain boundary, potential disorder, and surface scattering do not limit the mean free path; ferromagnetism with very strong uniaxial anisotropy, to favor moment flipping rather than canting as H is increased; and very strong spin-orbit coupling, magnifying the scattering cross-section of ``misaligned" spins. Transition metal dichalcogenides intercalated with various amounts of magnetic metals are promising materials where these optimal intercalation conditions may be achieved to maximize the observed MR, and such studies are currently underway.

\FloatBarrier

\section{Acknowledgements}
 W.J.H. H. J., and D.N. acknowledge support from DOE BES award DE-FG02-06ER46337. E.M., C.W.C. and A.M. acknowledge support by DOD PECASE.  J.S. acknowledges support by the Alexander von Humboldt Foundation.   We thank Dr. Wenhua Guo for performing electron diffraction.

\end{document}